\documentclass[12pt]{iopart}

\usepackage{graphicx}
  \expandafter\let\csname equation*\endcsname\relax
  \expandafter\let\csname endequation*\endcsname\relax
\usepackage{amsmath}
\usepackage{amsfonts}
\usepackage{bm}
\usepackage{color}

\newcommand{\ket}[1]{\ensuremath{\left|{#1}\right\rangle}}
\newcommand{\bra}[1]{\ensuremath{\left\langle{#1}\right |}}
\newcommand{\oper}[1]{\mathbf{\mathsf{#1}}}

\newcommand{\brm}[1]{\ensuremath{\mathbf{#1}}}
\newcommand{\scalprod}[2]{\ensuremath{\left\langle{#1}|{#2}\right\rangle}}
\newcommand{\sinc}{\ensuremath{\mathrm{sinc}}}
\newcommand{\rmvect}[1]{\boldsymbol{\mathrm{#1}}}

\begin{document}


\title[Generalized Hermite-Gauss decomposition of the two-photon state]{Generalized Hermite-Gauss decomposition of the two-photon state produced by spontaneous parametric down-conversion}

\author{S. P. Walborn and A. H. Pimentel}
\address{Instituto de F\'{\i}sica, Universidade Federal do Rio de
Janeiro, Caixa Postal 68528, Rio de Janeiro, RJ 21941-972, Brazil}
\ead{swalborn@if.ufrj.br}

\begin{abstract}
We provide a general decomposition of the two-photon state produced by spontaneous parametric down-conversion in Hermite-Gaussian modes, in the case that the pump beam is described by a Hermite-Gaussian beam of any order.  We show that the spatial correlations depend explicitly on the order of the pump beam, as well as other experimental parameters.  We use the decomposition to demonstrate a few interesting cases.  Our results are applicable to the engineering of two-photon spatial entanglement, in particular for non-Gaussian states.   
\end{abstract}



\maketitle
\section{Introduction}
The spatial correlations of photon pairs have been extensively studied in the context of quantum optics and quantum information \cite{walborn10}.  In particular, spatial correlations have played an important role in the understanding of fundamental quantum phenomena such as non-local interference \cite{fonseca99b,peeters09} and entanglement \cite{dangelo04,howell04,fedorov07,tasca08,gomes09b,pires09}.  Moreover, these spatial correlations may be useful for applications such as quantum imaging \cite{strekalov95,monken98a,fonseca99b,abouraddy01,abouraddy04,santos08} and cryptography \cite{almeida05,zhang08}.   For most investigations, correlated photon pairs are produced via Spontaneous Parametric Down Conversion (SPDC) \cite{walborn10}, which has served as a robust and controllable entanglement source. 
\par   
It is possible to explore continuous variable (CV) aspects of spatial correlations, and in this case one typically considers the near and far-field variables of some reference plane.  In this CV context, entanglement and non-locality have been observed using second-order Gaussian criteria \cite{dangelo04,howell04,tasca08,tasca09a}, as well as higher-order criteria that are more successful for certain non-Gaussian states \cite{gomes09b,walborn11a}.  It has also been shown that basic CV quantum gates can be implemented using spatial variables \cite{tasca11}.
CV spatial correlations can also be used to produce entangled $D$-dimensional systems \cite{neves05,hale05}.
\par   
One can also decompose the fields into a discrete set of transverse modes, such as the Laguerre-Gauss (LG) \cite{mair01,franke-arnold02,torres03a,torres03b,walborn04a,leach10,miatto11,miatto12b,salakhutdinov12} or Hermite-Gauss (HG) modes \cite{walborn05b,ren04,vanexter06,walborn07c,straupe11,miatto12a}.  Using the LG modes, the majority of experiments have been concerned with entanglement in orbital angular momentum \cite{mair01,walborn04a,oemrawsingh05,vanexter06,pires10}.  In terms of the HG modes, entanglement in the mode parity has been exploited for experimental violation of Bell's inequality \cite{yarnall07a}.   Propagation of entangled transverse modes through atmospheric turbulence \cite{pors11} and optical fibers has been studied \cite{loffler11}.  Investigations of quantum information protocols have relied on the entanglement in transverse modes \cite{vaziri02,langford04,barreiro08}.  To perform projective measurements in transverse modes, phase masks or plane-wave holograms, along with single mode fibers, are typically used.  For the HG modes, the phase masks introduce a phase shift according to $\mathrm{arg}[H_{n}(x)H_m(y)]$, where $H_n$ is the Hermite polynomial of order $n$ \cite{straupe11}.  Images of the first-order plane wave holograms are shown in Ref. \cite{langford04}.  Both of these devices transform a higher-order HG mode into the fundamental Gaussian mode, which can be selected using a single-mode optical fiber.  
\par
An expansion of the two-photon state into a basis of orthogonal transverse modes can be written as
\begin{equation}
\ket{\psi} = \sum_{j,k,s,t} C_{jkst} \ket{jk;\sigma_1}_1 \ket{st;\sigma_2}_2
\end{equation}
where  \ket{jk;\sigma} is single-photon in some transverse mode described by discrete indices $j,k$.  Here $\sigma_1$ and $\sigma_2$ describe the widths of the transverse modes.  The orthogonality condition is 
\begin{equation}
\langle{ab;\sigma}\ket{a^\prime b^\prime;\sigma} = \delta_{a,a^\prime}\delta_{b,b^\prime}.
\end{equation}  
Experimentally, one is free to choose the values of $\sigma_1$ and $\sigma_2$ by appropriate coupling into an optical system, such as a single mode fiber \cite{mair01,vaziri02,straupe11}.   An experimental study of this point has been reported in Ref. \cite{salakhutdinov12}.  
In the case of a Gaussian pump beam, under certain conditions this expansion has been studied theoretically for the Laguerre Gaussian modes \cite{franke-arnold02,torres03b,vanexter06,miatto11,miatto12b} and Hermite-Gaussian modes \cite{ren05,straupe11,miatto12a}.  Refs. \cite{walborn05b,walborn07c} considered higher-order Hermite-Gaussian pump beams under the condition that $\sigma_1=\sigma_2=\sqrt{2}w_p$, where $w_p$ is the radius of the Hermite-Gaussian pump beam.    
\par
Here we derive the expansion of the two-photon state produced by SPDC for Hermite-Gaussian pump beams of arbitrary order, using the Gaussian approximation of the phase matching function \cite{law04,chan07,miatto12b}.     In sections \ref{sec:state} and \ref{sec:gen} we review the two-photon state and the expansion in HG modes.  Explicit calculations are provided in the appendix.  In section \ref{sec:cases} we discuss our results and some interesting examples.   We show that, under certain conditions, the two-photon state is similar to a classical optical beam, and/or photon number states at the output of a beam splitter.   Our results are applicable to the engineering of two-photon non-Gaussian quantum states \cite{gomes09b,walborn11a}.     
\section{Two-photon state}
\label{sec:state}
Here we will be concerned with spatial correlations only. We thus assume that the pump laser and down-converted fields are monochromatic, paraxial and have well-defined polarizations.  These assumptions are justified experimentally by the crystal type and geometry, and by using narrow-band interference filters and polarizers, if necessary.  
For a sufficiently weak cw laser, the two-photon quantum state generated by
SPDC is \cite{walborn10,monken98a}
\begin{equation}
\ket{\psi}=\int\hspace{-2mm}\int\hspace{-1mm} d\brm{q}_{1}
d\brm{q}_{2}\ \Phi(\brm{q}_{1},\brm{q}_{2})\ket{\brm{q}_{1}}
\ket{\brm{q}_{2}}.
\label{eq:state}
\end{equation}
 The ket $\ket{\brm{q}_{j}}$ describes  a single-photon in a plane wave mode. The vector $\brm{q}=(q_x,q_y)$ is the transverse component of the wave vector
$\brm{k}$. The normalized angular spectrum of the two-photon state  $\Phi(\brm{q}_{1},\brm{q}_{2})$ is given by 
\begin{equation}
\Phi(\brm{q}_{1},\brm{q}_{2}) =\frac{1}{\pi}\sqrt{\frac{2L}{K}}\
\mathcal{V}(\brm{q}_{1}+\brm{q}_{2})\
\sinc\left(\frac{L|\brm{q}_{1}-\brm{q}_{2}|^{2}}{4K} \right),
\label{eq:state2}
\end{equation}
where $\mathcal{V}(\brm{q})$ is the normalized angular spectrum of the pump beam, $L$ is the length of the nonlinear crystal and $K$ is the wave number of the pump field.  
\par
 We now consider that the non-linear crystal is pumped with a Hermite-Gaussian beam $HG_{nm}$, generating a two-photon state \ket{\psi_{nm}}. 
We will denote $\mathcal{V}_{nm}(\rmvect{q})$ as the normalized two-dimensional angular spectrum of the HG mode.  The HG modes are separable in $x$ and $y$ coordinates, so that $\mathcal{V}_{nm}(q_x,q_y)=v_n(q_x)v_m(q_y)$, where
   \begin{equation}
v_{n}(q;w) =  \sqrt{w} D_{n}H_{n}\left(w q\right) 
 \exp\left(-\frac{w^2 q^2}{2}
\right) 
\label{eq:hgspdc6}
\end{equation}
and
\begin{equation}
D_{n} =  -i^{n} \sqrt{\frac{1}{2 ^{(n+1)} \sqrt{\pi} n!}}.
\label{eq:Dnm}
\end{equation}
 Here $H_{n}(x)$ is the $n^{\mathrm{th}}$-order Hermite polynomial. We have included the width parameter $w$ explicitly in the definition of the one-dimensional HG modes. 
Also interesting are the diagonal HG modes \cite{beijersbergen93}, which can be defined as
\begin{equation}
v_{n}(q_+;w) v_m(q_-;w) =\sum\limits_{k=0}^{n+m}B(n,m,k)v_{n+m-k}(q_x;w)v_k(q_y;w),
\label{eq:DHGexpan}
\end{equation}
where $q_{\pm}=q_x \pm q_y$ and
\begin{equation}
B(n,m,k) = \sqrt{\frac{(n+m-k)!k!}{2^{(n+m)}n!m!}}\frac{1}{k!}\frac{d^{k}}{dt^{k}}\left[(1-t)^{n}(1+t)^{m} \right]\big|_{t=0}.
 \label{eq:B}
\end{equation}

\section{Hermite-Gaussian Mode Decomposition}
\label{sec:gen}
 \par
 Since the HG beams form a complete basis, we can expand the two-photon state \eqref{eq:state} as
\begin{equation}
\ket{\psi_{nm}} = \sum_{j,k,s,t=0}^{\infty}C^{(nm)}_{jkst}\ket{{jk;\sigma_1}}\ket{{st;\sigma_2}},
\label{eq:stateHG}
\end{equation}
where we define
\begin{equation}
\ket{{\alpha\beta;\sigma}} = \int d\rmvect{q} v_{\alpha}(q_x;\sigma)v_{\beta}(q_y;\sigma) \ket{\rmvect{q}}.
\label{eq:vket}
\end{equation}
The expansion coefficients are given by the scalar product 
\begin{equation}
C^{(nm)}_{jkst} = \bra{{st;\sigma_2}}\scalprod{{jk;\sigma_1}}{\psi_{nm}}.
\label{eq:Cjkst}
\end{equation}

In Ref. \cite{walborn05b}, explicit analytical expressions for the coefficients $C^{(nm)}_{jkst}$ were derived for the two-photon state of the form \eqref{eq:state} with angular spectrum \eqref{eq:state2}.  There, it was considered that the width of the down-converted modes was $\sigma_1=\sigma_2=\sqrt{2}w_{p}$, where $w$ is the width of the pump field.  In Ref. \cite{ren05}, it was shown that for a thin crystal and a plane wave pump beam $\mathcal{V}(\brm{q}) \approx \delta(\brm{q})$, the down-converted photons are highly correlated in transverse modes, such that $C^{(00)}_{jkst} \approx \delta_{j,s} \delta_{k,t}$.  
\par
In many experimental conditions, the sinc function in \eqref{eq:state2} can be approximated by a Gaussian function, such that
\begin{equation}
\frac{1}{\pi}\sqrt{\frac{2L}{K}}\
\sinc\left(\frac{L|\brm{q}_{1}-\brm{q}_{2}|^{2}}{4K} \right) \approx \mathcal{V}_{00}(\brm{q}_{1}-\brm{q}_{2};\delta),
\end{equation}
where $\delta \approx 0.257 \sqrt{L/4K}$, so that the two functions have the same width at half maximum \cite{chan07}.  The validity of the Gaussian approximation has been discussed in Ref. \cite{gomez12,miatto12b}.
In this Gaussian approximation, the two-photon angular spectrum is separable in the $x$ and $y$ coordinates, and can be written as
\begin{align}
\Phi(\brm{q}_{1},\brm{q}_{2}) = & v_n({q}_{1x}+{q}_{2x}; w_x)v_m({q}_{1y}+{q}_{2y};w_y)\times \nonumber \\
& v_0({q}_{1x}-{q}_{2x};\delta_x)v_0({q}_{1y}-{q}_{2y};\delta_y),
\label{eq:state2}
\end{align}
\par
Likewise, the HG expansion coefficients can be separated as $C^{(nm)}_{jkst}=C^{(n)}_{js}C^{(m)}_{kt}$.   In Ref. \cite{straupe11}, an analytic expression was derived for the case of a Gaussian pump beam when the down-converted modes are characterized by widths $\sigma=\sigma^\prime=\sqrt{w \delta/2}$.  \par
Here we consider the general situation in which the down-converted modes are described by arbitrary width $\sigma$, as well as the case of higher-order Hermite-Gaussian pump beams $\mathcal{V}_{nm}$.   
  From these, one can construct the coefficients complete $C_{jkst}^{(nm)}$.  In particular, for $\sigma_1=\sigma_2=\sigma$, we find
\begin{align}
C_{ab}^{(n)} = & \sum\limits_{k=0}^{N} B(a,b,k) I(w_p,\sigma,n,N-k)I(\delta,\sigma,0,k),
\end{align}
where $B(a,b,k)$ is a coefficient defined in Eq. \eqref{eq:B} and 
\begin{equation}
I(\gamma,\sigma,t,s) = \int\hspace{-1mm}d{q} v_t(\sqrt{2} q;\gamma)  v^*_{s}(q;\sigma). 
\end{equation}
In the appendix, we derive an explicit expression for the coefficients $C^{(n)}_{ab}$.  Our results confirm earlier results that the parity is conserved in SPDC \cite{walborn05b,walborn07c,straupe11}, so that $\mathrm{parity}(a+b)=\mathrm{parity}(n)$.  Below, we provide some numerical examples and discuss some interesting cases. 
\section{Interesting Cases}
\label{sec:cases}
\subsection{Two-mode squeezed state}
By adjusting the width of the collected down-converted modes, one can manipulate the terms in the expansion \eqref{eq:stateHG}.  A particularly interesting case is when $\sigma\equiv \sigma^\prime=\sqrt{2 w \delta}$ and the pump beam is a Gaussian.  Here the coefficients are given by 
\begin{equation}
C^{(0)}_{js} =  \delta_{js} \frac{w \delta}{2} \frac{(w-\delta)^{j}}{(w+\delta)^{(j+1)}},
\label{eq:Cgauss}
\end{equation}  
where $\delta_{js}$ is the Kronecker delta.  
This corresponds to the Schmidt decomposition calculated and observed in Ref. \cite{straupe11}:
\begin{equation}
\ket{\psi_{00}} =  \sum_{ab=0}^{\infty} \sqrt{\lambda_{a}\lambda_b} \ket{ab;\sigma^\prime} \ket{ab;\sigma^\prime},
\label{eq:Gstate}
\end{equation}
where $\sqrt{\lambda_a}=C_{aa}$ and
we specify explicitly that the down-converted modes have widths $\sigma^\prime$.
This is similar to the two-mode squeezed vacuum state for intense fields, which displays photon number correlations between the modes \cite{braunstein05}.  Here, the correlation is between the HG mode indices in both the $x$ and $y$ directions.  For $w=\delta$, only one coefficient $C^{(00)}_{ab}=C^{(00)}_{00}=\lambda_0$ is nonzero, and the state is separable. This is true only in the Gaussian approximation.  However, for the wave function given in Eq. \eqref{eq:state2}, the state displays a minimum amount of entanglement \cite{law04,walborn07c}. 
\par
\begin{figure}
  \centering 
 \includegraphics[width=8cm]{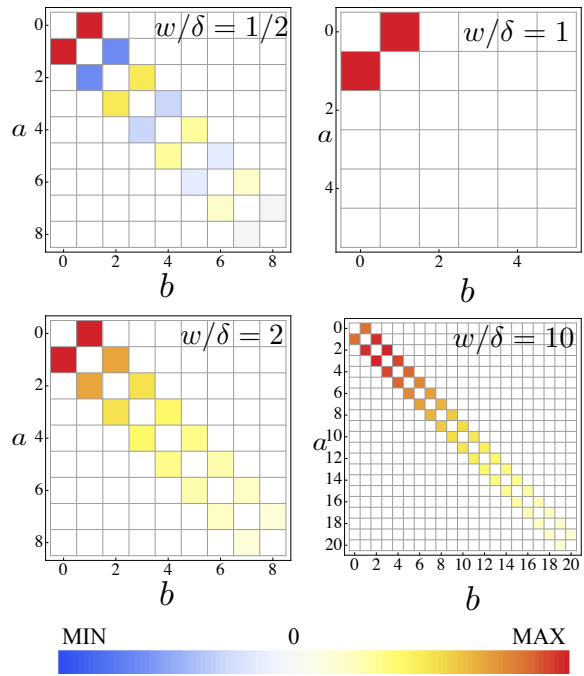} 
  \caption{The coefficients $C_{ab}^{(1)}$ for several values of $w/\delta$, with $\sigma=\sqrt{2 w \delta}$. The intensity scale goes from the largest negative (blue) value to the largest positive value (red).}
\label{fig:coeff1}
\end{figure}
\begin{figure}
  \centering 
 \includegraphics[width=8cm]{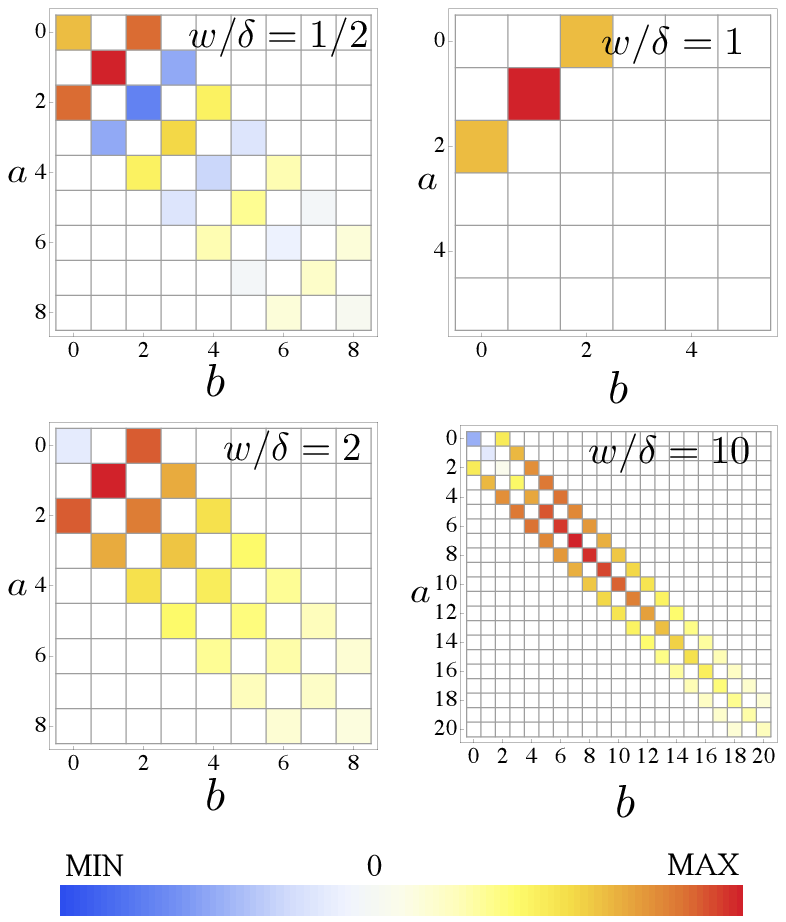} 
  \caption{The coefficients $C_{ab}^{(2)}$ for several values of $w/\delta$, with $\sigma=\sqrt{2 w \delta}$. The intensity scale goes from the largest negative (blue) value to the largest positive value (red).}
\label{fig:coeff2}
\end{figure}
\par
\begin{figure}
  \centering 
 \includegraphics[width=8cm]{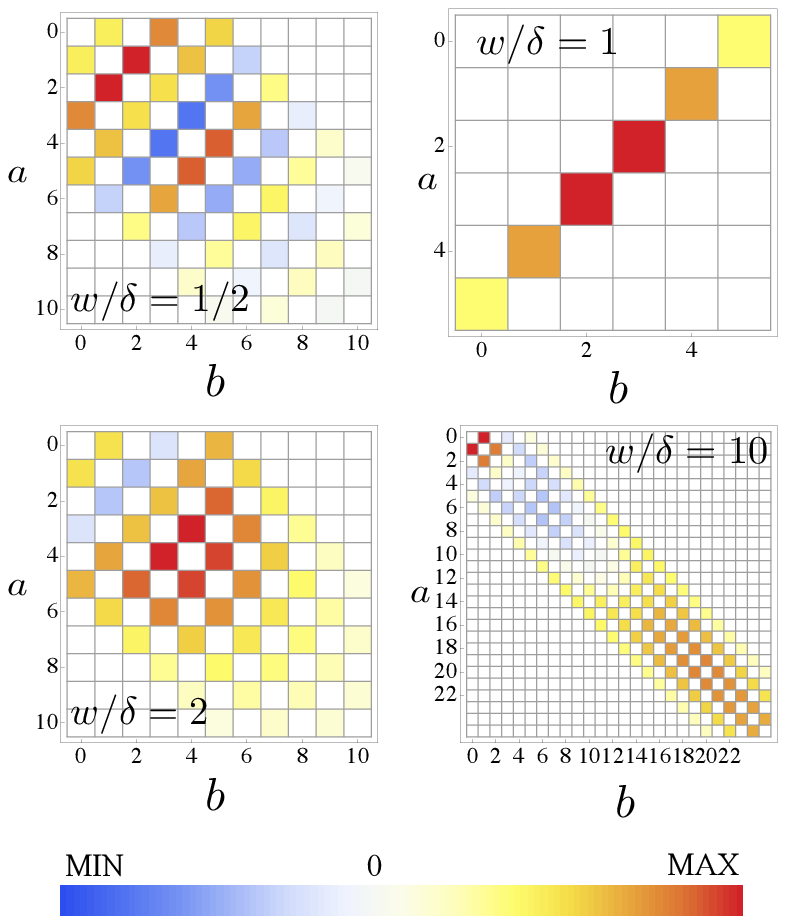} 
  \caption{The coefficients $C_{ab}^{(5)}$ for several values of $w/\delta$, with $\sigma=\sqrt{2 w \delta}$. The intensity scale goes from the largest negative (blue) value to the largest positive value (red).}
\label{fig:coeff5}
\end{figure}
\subsection{Correlated HG states}
  For the same width parameter $\sigma^\prime$, and considering now higher-order HG pump beams, the two-photon state takes the simplified form $\ket{\psi}_{nm}=\ket{\phi_n}\ket{\phi_m}$, where
\begin{equation}
\ket{\phi_{n}} =  \sum\limits_{\alpha=0}\limits^{\infty} \sum\limits_{r=0,\mathrm{par}(n)}^{n} C^{(n)}_{\alpha,\alpha+r} \left( \ket{\alpha}_1 \ket{\alpha+r}_2 +  \ket{\alpha+r}_1 \ket{\alpha}_2\right ), 
\label{eq:nonGstate}
\end{equation}
and similarly for $\ket{\phi_m}$. Here $\ket{\alpha}$ are single photons in one-dimensional HG modes, in analogy to Eq. \eqref{eq:vket}.  For $n,m\neq 0$, these states are always entangled.  
Figures \ref{fig:coeff1}, \ref{fig:coeff2} and \ref{fig:coeff5} show the coefficients $C^{(n)}_{\alpha\beta}$ for  various values of $\sigma^\prime$ in the cases $n=1,2,5$, respectively.   Different plots in each figure correspond to different values of $w/\delta$, the ratio between the width of the pump beam and the phase matching  function.  Correlation of the form given by Eq. \eqref{eq:nonGstate} can be observed.  For $w/\delta$, the states show the minimal amount of entanglement for each value of $n$, as has been shown in other studies \cite{law04}.  In addition, the minimal number of HG modes in the expansion increases with $n$, indicating that entanglement also increases with $n$ \cite{walborn07c}.    We note the alternating phase changes along the diagonal of the figures for coefficients of the HG decomposition when $w/\delta < 1$.  In the case of the Gaussian state described in Eqs. \eqref{eq:Cgauss} and \eqref{eq:Gstate}, the ratio $w/\delta$ determines whether the Gaussian functions of two-photon state presents momentum correlations ($\delta > w$, $q_1\approx q_2$) or anti-correlations ($\delta < w$, $q_1\approx - q_2$).  As can be seen in Eq. \eqref{eq:Cgauss}, when $j$ is odd the sign of the coefficient $C_{jj}^{(0)}$ is negative for correlations and positive for anti-correlations.  Thus, for $\delta > w$ the phase oscillates as a function of the sign of $j$, and this is related to the spatial symmetry (correlations or anti-correlations) of the two-photon state.  For higher order HG pump beams as in Figs. \ref{fig:coeff1} - \ref{fig:coeff5}, a similar effect is present.  In addition, the number of modes present depends explicitly on $|w/\delta|$, and the coefficient $C_{ab}^{(n)}$ decays exponentially with $a$ and $b$.   Thus, by controlling the ratio $|w/\delta|$, the number of transverse spatial modes can be manipulated.    This leads directly to control of the entanglement of the two-photon state \cite{law04,walborn07c,miatto12b}.  The states \eqref{eq:nonGstate} are similar to non-Gaussian states studied in Refs. \cite{gomes09a,gomes09b,walborn11a} with $n=1$, which display interesting properties.  We expect states with $n>1$ to display additional novel features \cite{gomes11}.       
\par
\begin{figure}
  \centering 
 \includegraphics[width=8cm]{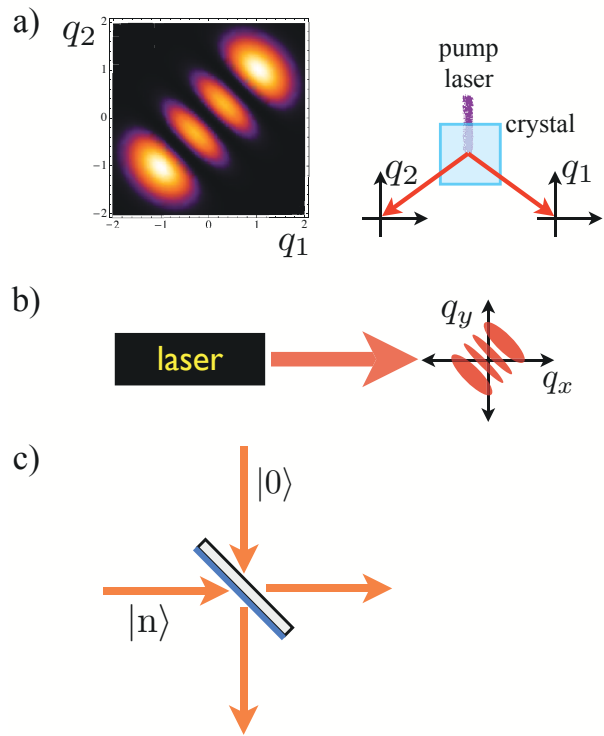} 
  \caption{ a) Left:  The probability distribution $|\phi_3|^2$ given by Eq. \eqref{eq:nonGstate}.  The two dimensional coordinate system cooresponds to one spatial dimension $q_1$ and $q_2$ of each photon produced by the non-linear crystal (right).  b) Analogy of state $\ket{\phi_n}$, given by Eq.  \eqref{eq:nonGstate}, with a classical Hermite-Gaussian mode with spatial variables $q_x$ and $q_y$.  c) The output state produced when a photon number state $\ket{\mathrm{n}}$ is superposed with vacuum $\ket{0}$ at a beam splitter is also analogous to two-photon state  \eqref{eq:nonGstate}.}
\label{fig:corr_beam}
\end{figure}
For $w_p=\delta$ and $\sigma=\sqrt{2w_p\delta}$, the coefficients $C_{ab}^{(n)}=\sqrt{n!/2^na!b!}$, and 
\begin{equation}
\ket{\phi_{n}} =  \sum\limits_{a=0}\limits^{n}  \sqrt{\frac{n!}{2^n a!(n-a)!}} \ket{n-a}_1 \ket{a}_2. 
\label{eq:nonGstate2}
\end{equation}
The quantum state describing the spatial variables of the photon pair are analogous to other physical situations in classical and quantum optics, as illustrated in Figure \ref{fig:corr_beam}.
  The states $\ket{\phi_n}$ in \eqref{eq:nonGstate2} are then analogous to the usual diagonal HG beams in classical optics \cite{beijersbergen93} defined in Eq. \eqref{eq:DHGexpan}, with index $m=0$.  In this case, it is straightforward to see that 
  \begin{equation}
 B(n,0,k) =  \sqrt{\frac{n!}{2^n n!(n-k)!}},
  \end{equation}
  giving 
  \begin{align}
v_{n}(q_+;w) v_0(q_-;w) =\sum\limits_{k=0}^{n}& \sqrt{\frac{n!}{2^n n!(n-k)!}} \nonumber \times \\
& v_{n-k}(q_x;w)v_k(q_y;w).
\label{eq:DHGexpan0}
\end{align}
The two-photon quantum state \eqref{eq:nonGstate2} is thus analogous to the two-dimensional classical beam defined in Eq. \eqref{eq:DHGexpan0},
  where the one transverse spatial variable of each photon--say $q_1$ and $q_2$--play the role of the transverse wave vectors $q_x$ and $q_y$ of the beam.  Similar two-photon states with this property have been observed for LG modes \cite{gomes09a,gomes11}.    
  \par  
To make an analogy with quantum optics of field modes, the wave function of state $\ket{\phi_n}$ is equivalent to combining a photon number state $\ket{\mathrm{n}}$ and a vacuum state on a 50/50 asymmetric beam splitter. The quantum state of the two modes at the output ports of the BS is
\begin{equation}
\frac{(\oper{a}_1^{\dagger}+\oper{a}_2^\dagger)^n}{\sqrt{2^n n!}}\ket{0}_1\ket{0}_2 = \sum\limits_{j=0}^n\sqrt{\frac{n!}{{2^n j!(n-j)!}}}\ket{n-j}_1\ket{j}_2,
\end{equation}
where $\oper{a}^{\dagger}$ is the usual creation operator.  Comparison with Eq. \eqref{eq:nonGstate2} shows that the two states are equivalent.  
\par
\begin{figure}
  \centering 
 \includegraphics[width=8cm]{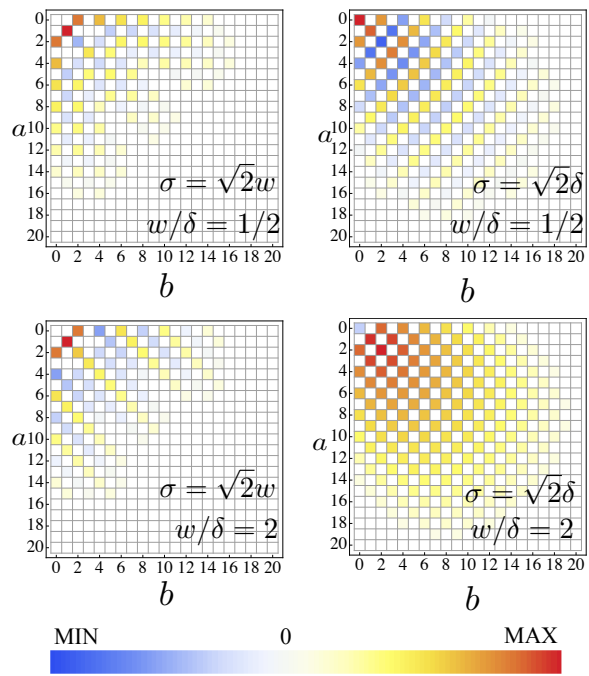} 
  \caption{The coefficients $C_{ab}^{(2)}$ for $\sigma=\sqrt{2} w$ and $\sigma=\sqrt{2} \delta$, with $w/\delta=2$ and $w/\delta=1/2$. The intensity scale goes from the largest negative (blue) value to the largest positive value (red).}
\label{fig:coeffgen}
\end{figure}

For the general case in which $\sigma \neq \sigma^\prime$, the quantum state still presents parity correlations.  Figure \ref{fig:coeffgen} shows four examples for $n=2$.   
If one chooses $\sigma=\sqrt{2}w_p$ (graphics on left), in addition to the parity restriction we have $a+b \geq n$ as can be seen by the absence of the $a=b=0$ term in the two graphics on the left, in accordance with previous work \cite{walborn05b,walborn07c}.  Thus, it is possible to produce two-photon states where lower order transverse modes are prohibited.  The two figures on the right correspond to $\sigma=\sqrt{2} \delta$. In this case parity correlations can also be observed, but we notice that there is also a nonzero coefficient when $a=b=0$.  Interestingly, when $w/\delta=2$, only the $a=b=0$ coefficient is negative.  For larger $w/\delta$, we observed numerically that additional negative coefficients appear.   
\par
\section{Conclusion}
 We have derived a Hermite-Gaussian expansion of the two-photon state produced by spontaneous parametric down-conversion when the pump beam is an arbitrary Hermite-Gaussian mode.  Use of the Gaussian approximation to the phase matching function has allowed us to arrive at analytical expressions for the expansion coefficients.  Our results show that a number of interesting two-photon states can be produced by changing the Hermite-Gauss indices of the pump beam, as well as the widths of the pump beam and the phase matching function.  An interesting feature of the HG mode expansion is the fact that, under certain conditions, one can consider a single spatial dimension for each photon.  This allows for simplification of the theoretical analysis and experimental investigations \cite{straupe11}.  We have also shown that in certain cases the two-photon state in one spatial dimension is analogous to a diagonal Hermite-Gaussian beam in classical optics, and also to a photon number state superposed with vacuum at a beam splitter.    The analogy between a single spatial dimension and a single harmonic oscillator allows for interesting parallels to be drawn.  For instance, viewing the spatial correlations as forming a "two-photon beam", as  illustrated in Fig. \ref{fig:corr_beam} a), allows one to identify interesting features, such as the correlation vortex reported in Refs. \cite{gomes09a,gomes11}.  Since this image relies on a single spatial dimension of each photon, it is more adequately described using the HG modes, described in cartesian coordinate systems.     We expect our results to be useful in the engineering of spatial correlations, in particular in the case where non-Gaussian properties of the quantum state are to be explored \cite{gomes09b,walborn11a,gomes09a}.       
\ack
Financial support was provided by Brazilian agencies CNPq, 
CAPES, FAPERJ, and the Instituto Nacional de Ci\^encia e Tecnologia (INCT) - Informa\c{c}\~ao Qu\^antica.  
\section*{Appendix}
The one-dimensional expansion coefficients are  $C^{(n)}_{ab}=\bra{v_{a}}\scalprod{v_{b}}{\psi_{n}}$.  Using the definition of HG modes \eqref{eq:hgspdc6} and the two-photon state \eqref{eq:state}, we have
\begin{align}
C_{ab}^{(n)} =  \int\hspace{-2mm}\int\hspace{-1mm}  & v_n(q_1+q_2;w_p) v_0(q_1-q_2;\delta) \times \nonumber \\
& v^*_a(q_1;\sigma_1) v^*_b(q_2;\sigma_2) d{q}_{1}
d{q}_{2}
\label{eq:int1}
\end{align}
Defining 
\begin{equation}
q_\pm = \frac{1}{\sqrt{2}} (q_1 \pm q_2)
\end{equation}
we can rewrite the integral in terms of $q_\pm$ variables.  Let us consider that the down-converted modes satisfy $\sigma_1=\sigma_2\equiv \sigma$.  Using the definition of the diagonal HG modes \eqref{eq:DHGexpan}, we can rewrite Eq. \eqref{eq:int1} as 
\begin{align}
C_{ab}^{(n)} = & \sum_{k=0}^N B(a,b,k) \int\hspace{-1mm}d{q}_{+}
 v_n(\sqrt{2} q_+;w_p)  v^*_{N-k}(q_+;\sigma)  \times \nonumber \\
 & \int\hspace{-1mm} d{q}_{-} v_0(\sqrt{2} q_-;\delta)  v^*_{k}(q_-;\sigma),
\label{eq:int2}
\end{align}
where $N=a+b$ and $B(a,b,k)$ is defined in Eq. \eqref{eq:B}. 
The integrals in Eq. \eqref{eq:int2} can be solved separately. Then coefficients can then be written as
\begin{align}
C_{ab}^{(n)} = & \sum\limits_{k=0}^{N} B(a,b,k) I(w_p,\sigma,n,N-k)I(\delta,\sigma,0,k),
\label{eq:coeff}
\end{align}
where 
\begin{equation}
I(\gamma,\sigma,t,s) = \int\hspace{-1mm}d{q} v_t(\sqrt{2} q;\gamma)  v^*_{s}(q;\sigma) 
\end{equation}
Expanding the Hermite polynomials in power series \cite{lebedev72} and integrating each term, one has
\begin{align}
I(\gamma,\sigma,t,s)
 & =  \frac{\gamma \sigma D_tD^*_s t! s!}{\gamma^2+\sigma^2/2}\sum\limits_{\ell=0}^{\mathrm{floor}(t/2)}\frac{(-1)^{\ell}}{\ell! (t-2\ell)!} \times \nonumber \\
& \sum\limits_{j=0}^{\mathrm{floor}(s/2)}\frac{(-1)^{j}}{j! (s-2j)!}  \times \nonumber \\
& \left(\frac{2\sqrt{2}\gamma}{\gamma^2+\sigma^2/2}\right)^{t-2\ell}\left(\frac{2\sigma}{\gamma^2+\sigma^2/2}\right)^{s-2j} \times \nonumber \\
& \Gamma\left(\frac{1}{2}+\frac{t}{2}+\frac{s}{2}-j-\ell \right),
\end{align}
where $\Gamma$ is the usual Gamma function.
\par
When $t=0$, $I(\gamma,\sigma,0,s)$ can be simplified to
\begin{equation}
I(\gamma,\sigma,0,s) =  \frac{\sqrt{\pi} \gamma \sigma D_0D^*_s s!}{\frac{s}{2}!(\gamma^2+\frac{\sigma^2}{2})}\left(\frac{\sigma}{\gamma^2+\frac{\sigma^2}{2}}-1 \right)^{s/2}.
\end{equation}

\bibliographystyle{vancouver}

\end{document}